\documentclass[english,aps,superscriptaddress,pra,amsmath,amssymb,preprint]{revtex4}
\usepackage[T1]{fontenc}
\usepackage[utf8]{luainputenc}
\usepackage{amsmath}
\usepackage{graphicx}
\usepackage{esint}

\makeatletter
\@ifundefined{textcolor}{}
{%
 \definecolor{BLACK}{gray}{0}
 \definecolor{WHITE}{gray}{1}
 \definecolor{RED}{rgb}{1,0,0}
 \definecolor{GREEN}{rgb}{0,1,0}
 \definecolor{BLUE}{rgb}{0,0,1}
 \definecolor{CYAN}{cmyk}{1,0,0,0}
 \definecolor{MAGENTA}{cmyk}{0,1,0,0}
 \definecolor{YELLOW}{cmyk}{0,0,1,0}
}

\makeatother

\usepackage{babel}
\begin{document}

\preprint{This line only printed with preprint option}

\title{What is the fate of runaway positrons in tokamaks?}

\author{Jian Liu}
\affiliation{Department of Modern Physics and Collaborative Innovation Center
for Advanced Fusion Energy and Plasma Sciences, University of Science
and Technology of China, Hefei, Anhui 230026, China}

\author{Hong Qin}
\email{hongqin@ustc.edu.cn}
\affiliation{Department of Modern Physics and Collaborative Innovation Center
for Advanced Fusion Energy and Plasma Sciences, University of Science
and Technology of China, Hefei, Anhui 230026, China}
\affiliation{Princeton Plasma Physics Laboratory, Princeton University, Princeton,
NJ 08543}

\author{Nathaniel J. Fisch}
\affiliation{Princeton Plasma Physics Laboratory, Princeton University, Princeton,
NJ 08543}

\author{Qian Teng}
\affiliation{Princeton Plasma Physics Laboratory, Princeton University, Princeton,
NJ 08543}

\author{Xiaogang Wang}
\affiliation{School of Physics, Peking University, Beijing, China 100871}

\begin{abstract}
Massive runaway positrons are generated by runaway electrons in tokamaks.
The fate of these positrons encodes valuable information about the
runaway dynamics. The phase space dynamics of a runaway position is
investigated using a Lagrangian that incorporates the tokamak geometry,
loop voltage, radiation and collisional effects. It is found numerically
that runaway positrons will drift out of the plasma to annihilate
on the first wall, with an in-plasma annihilation possibility less
than 0.1\%. The dynamics of runaway positrons provides signatures
that can be observed as diagnostic tools.
\end{abstract}
\maketitle
Positron, the anti-particle of electron, is a rare species in the
part of universe where we reside. Since last century, man-made positrons
have been generated in accelerators for scientific research, in nuclear
reactors as the byproducts, and applied in different fields, such
as medicine and material detection \cite{PETbook,1980AnRMS..10..393S,van1997testing}.
Recent researches indicate that tokamak, a magnetic confinement fusion
energy device, may be the largest artificial positron factory in the
world \cite{Posi_Creation,PhysRevLett.108.225003}. In large tokamaks
like JET and JT-60U, above $10^{14}$ positrons are generated in a
post-disruption plasma by runaway electrons \cite{Dreicer,Kulsrud73,connor1975relativistic,helander2002runaway,Knoepfel,Rosenbluth97,Bakhtiari,Run_dynamics,key-2,x_ray,key-5,syncrotron_radiation,neutron,bolt1987simulation,nygren1997runaway,bartels1994impact,kawamura1989generation}.
The dynamics of these positrons after birth in tokamaks is a noteworthy
question that may yield valuable information about the runaway dynamics
and disruption process in tokamaks. What is the fate of these positrons?
Will they annihilate inside the plasma or on the first wall of the
vacuum chamber? Because the annihilation probability dependents on
the path and velocity, the fate of runaway positrons is determined
by their phase space trajectories, which are strongly affected by
the loop electric field, the helical magnetic field, and the collisional
and radiation effects. Incorporating all these factors, we study the
phase space dynamics of runaway positrons in tokamaks and predict
their final fate.

\begin{figure}
\begin{centering}
\includegraphics[width=7cm]{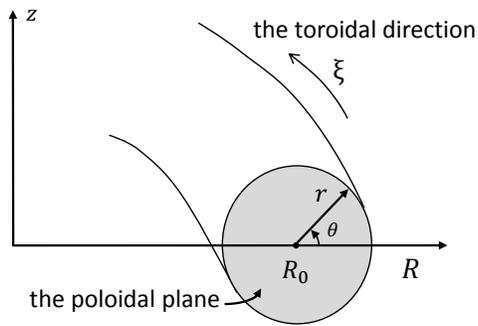}
\par\end{centering}

\caption{\label{fig:3D_sructure}The circular concentric magnetic flux surfaces
and the coordinate systems.}
\end{figure}

When relativistic runaway electrons interact with the thermal electrons
and ions, positrons are produced in the pair production process. In
tokamaks, because the energy of runaway electrons is typically above
10 MeV, most of their ``children'' positrons are born relativistic
and can be accelerated to runaway velocity by the loop voltage. In
the toroidal direction, the runaway positrons are accelerated along
the toroidal electric field and slowed down by radiation loss and
collisions. As the energy increase, strong synchrotron radiation and
bremsstrahlung radiation begin to dominate in the drag force, which
finally balances the loop electric field force. The collisional drag
from the background plasma becomes small after runaway positrons gain
high velocities. Nevertheless, the collisional effect offers a momentum
transfer mechanism between the parallel and the perpendicular momentum
through the pitch-angle scattering. On the other hand, the projection
of the positron gyrocenter trajectory onto the poloidal plane is not
exactly located on a flux surface due to the geometric effect of the
tokamak magnetic field. Numerical results reveal that the circular
orbits of runway positrons in the poloidal plane drift toward or against
the major radius direction $\hat{\boldsymbol{e}}_{R}$ (see Fig.\-\ref{fig:3D_sructure}),
under the influence of the loop electric field along the toroidal
direction $\hat{\xi}$ and the helical background magnetic field.
Due to the drift effect, the runaway positrons hit the first wall
of the tokamak within about one hundred milliseconds with energy as
large as $150$ MeV. After the phase space trajectories are known,
the annihilation probabilities along the trajectories can be calculated.
For a typical positron, it is found that the probability of annihilation
inside the plasma is only about 0.1\%. Essentially all positrons generated
in the tokamak will annihilate on the first wall of the vacuum chamber.
This fact suggests that the annihilation spectrum from the wall can
be analyzed to infer the dynamics and distribution of the runaway
positions. Admittedly, there are many other loss mechanisms for the
positions. For example, stochastic field lines \cite{yoshino2000runaway,helander2000suppression}
induced by MHD modes and ripple field \cite{0741-3335-35-SB-010}
can result in transport or orbit loss for energetic positrons. The
orbit loss considered in the present study is a neoclassical effect
due to the toroidal geometry, and is probably the slowest loss mechanism
among all possible mechanisms. The in-plasma annihilation probability
of 0.1\% obtained in our calculation is thus its upper bound.

We now present in detail the study of phase space dynamics of runaway
positrons. For theoretical analysis and numerical simulation, the
gyrocenter of runaway positrons is described by a Lagrangian, which
incorporates the tokamak geometry, loop voltage, radiation and collisional
effects. Parallel momentum, perpendicular momentum, annihilation rate,
and the drift orbit in the poloidal plane are numerically calculated
as functions of time. The potential of runaway positrons as a diagnostic
tool is discussed at the end.

As the anti-particle, the positron has the same rest mass as the electron,
denoted as $m_{e}$, but opposite electric charge, denoted as $e$.
So the dynamics of runaway positrons are similar to that of runaway
electrons. The Lorentz factor for a positron with momentum $p$ is

\begin{equation}
\gamma=\sqrt{1+\frac{p^{2}}{m_{e}^{2}c^{2}}}=\sqrt{1+\frac{p_{\parallel}^{2}}{m_{e}^{2}c^{2}}+\frac{p_{\perp}^{2}}{m_{e}^{2}c^{2}}}\,,
\end{equation}
where $c$ is the light speed in vacuum, $p_{\parallel}$ is the momentum
component parallel to the background magnetic field, and $p_{\perp}$
is the perpendicular momentum. Its synchrotron radiation drag force
takes the form
\begin{equation}
F_{s}=\frac{2}{3}r_{e}m_{e}c^{2}\gamma(\gamma^{2}-1)^{3/2}\left(\frac{1}{R_{0}^{2}}+\frac{\sin^{4}\theta}{r_{g}^{2}}\right)\,,
\end{equation}
where $r_{e}=e^{2}/4\pi\epsilon_{0}m_{e}c^{2}$ is the classical positron
radius, $r_{g}=p_{\perp}/eB$ is the positron gyro-radius, $\theta$
is the pitch angle defined by $\sin\theta=p_{\perp}/p$, and $R_{0}$
is the major radius of the tokamak. The bremsstrahlung drag force
is

\begin{equation}
F_{B}=\frac{4}{137}n_{e}m_{e}\gamma c^{2}r_{e}^{2}\left(Z_{eff}+1\right)\left(\ln2\gamma-\frac{1}{3}\right)\,,
\end{equation}
where $n_{e}$ is the number density of the background plasma, $Z_{eff}$
is the effective ion charge factor. The collisional friction force
is

\begin{equation}
F_{c}=\frac{n_{e}e^{4}m_{e}\ln\Lambda}{4\pi\epsilon_{0}^{2}}\frac{\gamma^{2}}{p^{2}}\,,
\end{equation}
where $\ln\Lambda$ is the Coulomb logarithm, which varies slowly
with the plasma parameters. To theoretically investigate the dynamics
of runaway positrons in tokamaks, we describe the dynamics of positrons
by the following Lagrangian \cite{grebogi1984relativistic},
\begin{equation}
L=(e\mathbf{\mathbf{A}_{0}}+e\mathbf{A}_{l}+e\mathbf{A}_{eff\parallel}+p_{\Vert}\mathbf{b})\cdot\dot{\mathbf{x}}-\gamma mc^{2}\,\text{.}
\end{equation}
Here, $\mathbf{A}_{0}$ is the vector potential of the background
magnetic field satisfying $\mathbf{B}=\nabla\times\mathbf{A}_{0}$,
$\mathbf{A}_{l}$ is the vector potential of the loop electric field
satisfying
\begin{equation}
-\frac{\partial\mathbf{A}_{l}}{\partial t}=\mathbf{E}_{loop}\,,
\end{equation}
$\mathbf{A}_{eff\parallel}$ is the parallel component of the effective
vector potential corresponding to the drag force,
\begin{equation}
\mathbf{A}_{eff\parallel}=\frac{p_{\parallel}}{p}\frac{t}{e}\left(F_{s}+F_{B}+\frac{Z_{eff}+\gamma+1}{\gamma}F_{c}\right)\mathbf{b}\,,
\end{equation}
and $\mathbf{b}$ is the unit vector along the magnetic field. The
magnitude of the effective vector potential in the perpendicular direction
is
\begin{equation}
A_{eff\perp}=\frac{p_{\perp}}{p}\frac{t}{e}\left[F_{s}+F_{B}+\left(1-\frac{p_{\parallel}^{2}}{p_{\perp}^{2}}\frac{Z_{eff}+1}{\gamma}\right)F_{c}\right]\,.
\end{equation}

Without loss of generality, we consider a tokamak magnetic field with
circular concentric flux surfaces,

\begin{equation}
\mathbf{B}=\frac{B_{0}R_{0}}{R}\mathbf{\hat{e}}_{\xi}+\frac{B_{0}r}{qR}\mathbf{\hat{e}}_{\text{\ensuremath{\theta}}}\,.
\end{equation}
Accordingly we choose its vector potential to be
\begin{equation}
\mathbf{A}_{0}=\frac{B_{0}R_{0}z}{2R}\mathbf{\hat{e}}_{R}+\frac{B_{0}r^{2}}{2Rq}\mathbf{\hat{e}}_{\xi}+\frac{R_{0}B_{0}}{2}\ln\left(\frac{R}{R_{0}}\right)\mathbf{\hat{e}}_{z}\,.
\end{equation}
The loop voltage is set to be
\begin{equation}
\mathbf{E}_{loop}=E_{l}\frac{R_{0}}{R}\mathbf{\hat{e}}_{\xi}\,.
\end{equation}

\begin{figure}
\begin{centering}
\includegraphics[scale=0.75]{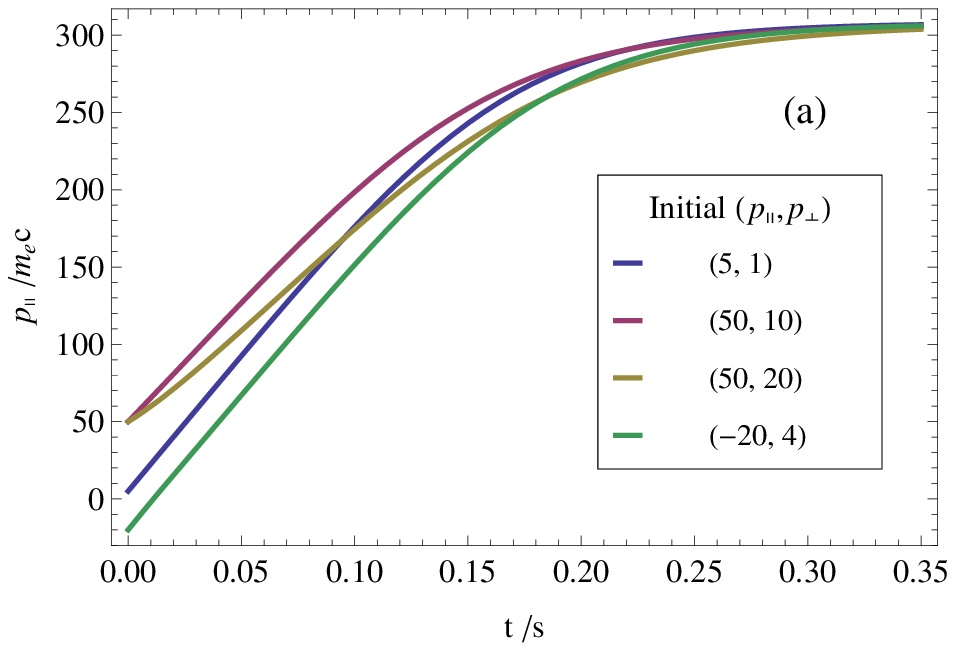}
\par\end{centering}

\begin{centering}
\includegraphics[scale=0.74]{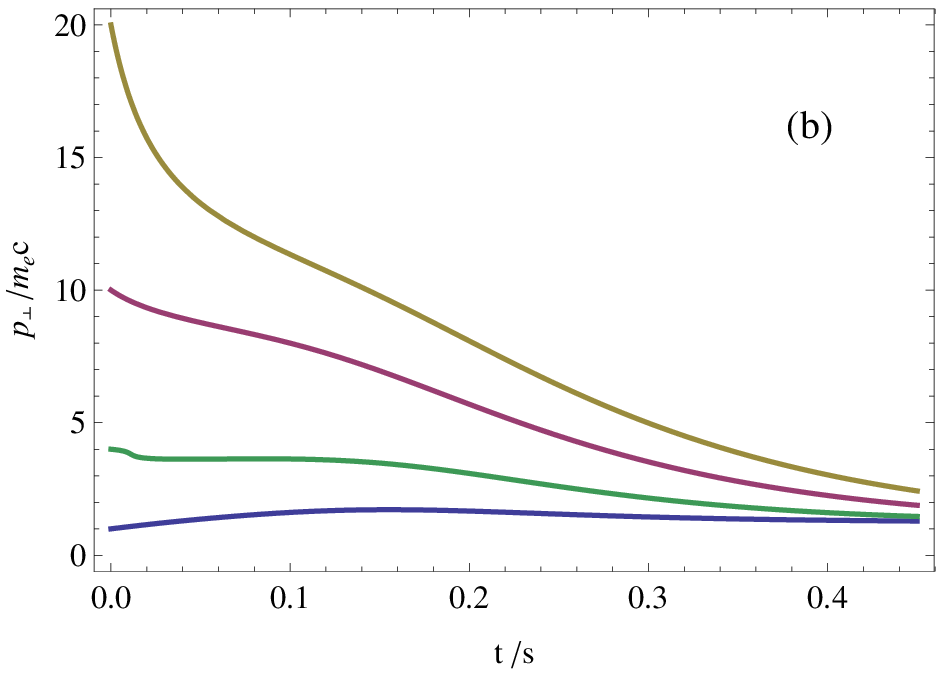}
\par\end{centering}

\begin{centering}
\includegraphics[scale=0.73]{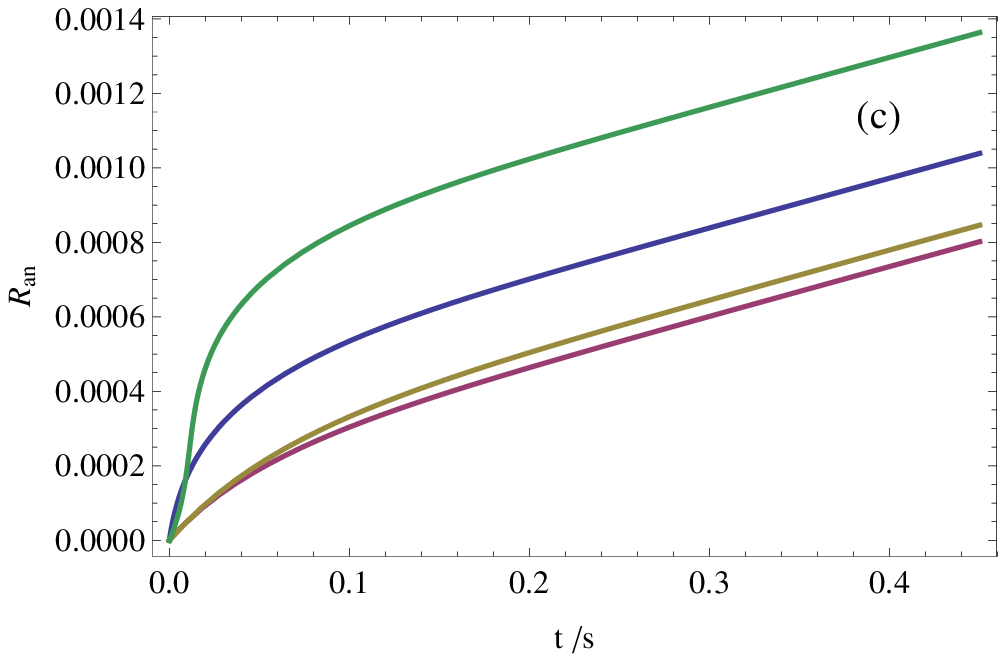}
\par\end{centering}

\caption{\label{fig:The-evolution-curves}The evolution of the parallel momentum
(a), the perpendicular momentum (b), and the annihilation rate (c)
of runaway positrons in tokamaks with different initial values of
$(p_{\parallel}\text{,}p_{\perp})$ (normalized by $m_{e}c$). The
loop electric field is $E_{l}=3V/m$.}
\end{figure}

\begin{figure}
\begin{centering}
\includegraphics[scale=0.37]{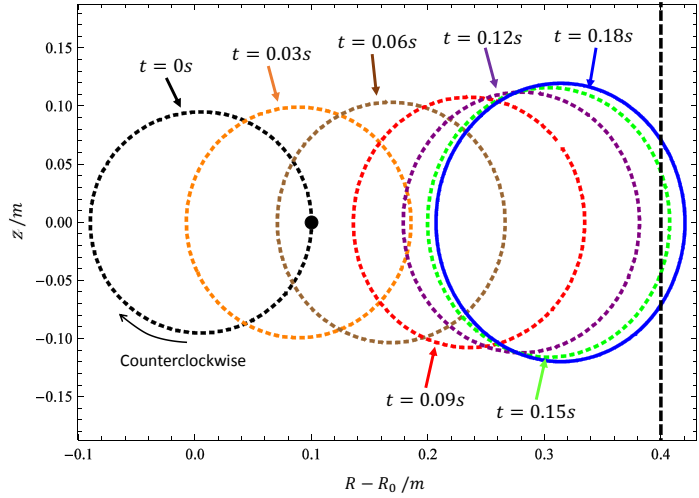}
\par\end{centering}

\caption{\label{fig:snapshot}Snapshots, taken for every $0.03s$, of the circular
orbit of a runaway positron in the poloidal plane, with initial position
$R-R_{0}=0.1m$, $z=0m$ and initial momenta $p_{\parallel}=5m_{e}c$
and $p_{\perp}=m_{e}c$. The loop electric field is $E_{l}=5V/m$.
The position of the first wall is indicated by the dashed vertical
line.}

\end{figure}

In the right-handed coordinate system $(R,\xi,z)$ (see Fig.\-\ref{fig:3D_sructure}),
the Lagrangian takes the form
\begin{equation}
L=p_{R}\dot{R}+p_{\xi}\dot{\xi}+p_{z}\dot{z}-\gamma m_{e}c^{2}\,,\label{eq:Lag2}
\end{equation}
where
\begin{equation}
p_{R}=\frac{eB_{0}R_{0}z}{2R}-\frac{B_{0}z}{BqR}\left(p_{\parallel}+eA_{eff\parallel}\right)\,,
\end{equation}
\begin{equation}
p_{\xi}=\frac{eB_{0}r^{2}}{2q}+eA_{l}R_{0}+\frac{B_{0}R_{0}}{B}\left(p_{\parallel}+eA_{eff\parallel}\right)\,,
\end{equation}
\begin{equation}
p_{z}=-\frac{eR_{0}B_{0}}{2}\ln\left(\frac{R}{R_{0}}\right)+\frac{B_{0}x}{BqR}\left(p_{\parallel}+eA_{eff\parallel}\right)\,.
\end{equation}
Because of the toroidal symmetry, i.e. $\partial L/\partial\xi=0$,
the effective toroidal momentum is conserved, i.e.,

\begin{equation}
p_{\xi}=\frac{\partial L}{\partial\dot{\xi}}=\mathrm{const}\,.
\end{equation}
This invariance determines the evolution of $p_{\parallel}$ as
\begin{equation}
p_{\parallel}=\frac{B}{B_{0}R_{0}}\left(p_{\xi}-\frac{eB_{0}r^{2}}{2q}-eA_{l}R_{0}\right)-eA_{eff\parallel}\,.\label{eq:p_para}
\end{equation}
Meanwhile, there exits another conserved quantity, the effective magnetic
moment $\mu$, defined by

\begin{equation}
\mu=\frac{(p_{\perp}+A_{eff\perp})^{2}}{2m_{e}B}\,,\label{eq:p_perp}
\end{equation}
which determines the evolution of $p_{\perp}$. If neglecting the
higher-order terms caused by the toroidal effect and the poloidal
field, Eqs.\-(\ref{eq:p_para}) and (\ref{eq:p_perp}) give

\[
\begin{aligned} & \frac{\text{\ensuremath{\mathrm{d}}}p_{\parallel}}{\mathrm{d}t}=eE_{loop}-\frac{p_{\parallel}}{p}\left(F_{s}+F_{B}+\frac{Z_{eff}+\gamma+1}{\gamma}F_{c}\right)\,,\\
 & \frac{\text{\ensuremath{\mathrm{d}}}p_{\perp}}{\mathrm{d}t}=-\frac{p_{\perp}}{p}\left[F_{s}+F_{B}+\left(1-\frac{p_{\parallel}^{2}}{p_{\perp}^{2}}\frac{Z_{eff}+1}{\gamma}\right)F_{c}\right]\,,
\end{aligned}
\]
which are consistent with the momentum evolution equations in \cite{Bakhtiari,key-4,martin1999energy,andersson2001damping}.
After substituting Eqs.\-(\ref{eq:p_para}) and (\ref{eq:p_perp})
into Eq.\-(\ref{eq:Lag2}) and dropping the term $p_{\xi}\dot{\xi}$,
the toroidal symmetry leads to the reduced Lagrangian in the $(R,z)$
2D space
\begin{equation}
L=p_{R}(R,z)\dot{R}+p_{z}(R,z)\dot{z}-H(R,z)\,.
\end{equation}
This is the procedure of Routh reduction. Then the runaway positron
dynamics in the 2D configuration space, i.e., the projection poloidal
plane, is given by the Euler-Lagrangian equation,

\begin{equation}
\dot{R}=\frac{\partial H/\partial z}{\partial p_{R}/\partial z-\partial p_{z}/\partial R}\,,\label{eq:Rdot}
\end{equation}
\begin{equation}
\dot{z}=\frac{\partial H/\partial R}{\partial p_{z}/\partial R-\partial p_{R}/\partial z}\,.\label{eq:zdot}
\end{equation}
 Equations (\ref{eq:p_para}), (\ref{eq:p_perp}), (\ref{eq:Rdot})
and (\ref{eq:zdot}) determines the dynamics of runway positrons in
phase space. Given the dynamics in the momentum space, the in-plasma
annihilation probability of a runaway positron can be calculated according
to
\begin{equation}
R_{an}=\intop_{0}^{t}n_{e}\sigma_{an}v\mathrm{d}\tau\,,
\end{equation}
where the annihilation cross-section for the positron-electron reaction
is \cite{PhysRevLett.108.225003}

\begin{equation}
\sigma_{an}=\frac{\pi r_{e}^{2}}{1+\gamma}\left[\frac{\gamma^{2}+4\gamma+1}{\gamma^{2}-1}\ln(\gamma+\sqrt{\gamma^{2}-1})-\frac{\gamma+3}{\sqrt{\gamma^{2}-1}}\right]\frac{2\pi}{137p(1-e^{-2\pi/137p})}\,.
\end{equation}

To guarantee the long term numerical accuracy and fidelity, we have
adopted a variational symplectic integrator \cite{qin2008variational,qin2009variational},
which discretizes the Lagrangian directly, to carry out the numerical
simulation. For conventional integration algorithms, such the Runge-Kutta
method, numerical errors from different time-steps accumulate coherently
and the error grows without bound for long-term simulations. In comparison,
the variational symplectic integrator can globally bound the numerical
error for all time-steps, and thus are superior for simulating the
runaway dynamics, which often involves many hundreds of thousands
of turns in the poloidal plane.

For the present study, we use the parameters from EAST as a specific
example \cite{wu2007overview}. We set $B_{0}=3.5T$, $R_{0}=1.7m$,
$n_{e}=10^{19}m^{-3}$, and $\ln\Lambda=10$. The loop electrical
field is $E_{l}=3V/m$, and the minor radius is $a=0.4m$. In typical
EAST operations, the safety factor $q$ varies in the range of $1$
to $3$, while the profile of $q$ is rather flat in the core region.
Thus we choose $q=2$ in the calculation. For runaway positrons with
different initial parallel and perpendicular momentum, their momentum
evolution are plotted in Fig.\-\ref{fig:The-evolution-curves}. After
$0.3s$, all the parallel momenta reach a steady value around $300m_{e}c$,
as a result of the balance between the loop electric field acceleration
and the drag force resistance. The perpendicular momenta also evolve
towards a steady value, though a little slower, due to the balance
between the radiation loss and the momentum transfer from parallel
direction through the pitch angle scattering. The green curves show
the dynamics of what might be called a \textquotedbl{}backward runaway\textquotedbl{}
positron \cite{karney1986current,fisch1987theory}, whose initial
parallel momentum is opposite to the loop electric field. It undergoes
a deceleration under in the toroidal direction at first. However,
its parallel momentum reverses sign after $0.01s$ and then increases
like forward runaway positrons with positive parallel momenta. Its
perpendicular momentum has a drop in the deceleration phase and passes
a point of inflexion at $0.01s$. It can be seen that runaway positrons
of widely different initial momentum nonetheless approach attractor
curves in momentum space. The existence of attractor curves is made
possible because of the dissipation introduced by radiation effects.

The possibility of annihilation for runaway positrons with different
initial momenta is depicted in Fig.\-\ref{fig:The-evolution-curves}(c).
Because the annihilation cross-section becomes very large when the
positron moves slowly relative to the background electrons, the annihilation
rate of the positron with negative parallel momentum grows rapidly
during its turning around process. Nevertheless, it escaped from the
doom of annihilation for a non-relativistic positron. Though its parallel
momentum goes through zero, its perpendicular momentum is still relativistic,
which ensures that it is not annihilated in the slowing-down process
and can be accelerated to runaway in the backward direction. Overall,
the annihilation probability is very low for runaway positrons. The
probability is only about $0.1\%$ after $0.4s$, which is long enough
for runaway positrons to escape the tokamak plasmas through the drift
motion or other transport/loss mechanisms in the poloidal plane.

The motion of a runaway positron in the poloidal plane with loop electric
field $E_{l}=5V/m$ is depicted in Fig.\-\ref{fig:snapshot}. The
loop electric field and the toroidal magnetic field are in the $\hat{\boldsymbol{e}}_{\xi}$
direction, and the poloidal magnetic field is in the $\hat{\boldsymbol{e}}_{\theta}$
direction. The counterclockwise circular orbits are snapshotted for
every $0.03$s. The outward drift of the circular orbit to the $\hat{\boldsymbol{e}}_{R}$
direction is evident. The drift velocity slows down with the radiation
resistance in the toroidal direction increases. The drift velocity
can be roughly estimated as $v_{dr}=q(E_{l}+E_{eff\parallel})/B_{0}$
\cite{Run_dynamics}. The runaway positron finally hits the wall before
$t=0.15s$, with an in-plasma annihilation probability less than $0.1\%$.
This example illustrates the fact that most of the runaway positrons
in tokamaks can drift out of the plasma to hit the wall before annihilation
within the plasma. If the safety factor $q=q(R,z)$ is a function
of the spatial coordinates like in real tokamaks, the drift velocity
will explicitly depend on the spatial location. However, the space-dependent
safety factor also leads to similar runaway positron dynamics and
the same main conclusions as the constant-$q$ assumption. For situations
with larger loop electric field, the runaway positrons will drift
faster. If changing the helical direction of the magnetic field or
the direction of the loop electric field, the positron will drift
inwards, i.e., towards the negative $\hat{\boldsymbol{e}}_{R}$ direction,
to hit the inner wall. It is observed that the radius of circular
orbit varies as the positron drifts along the $\hat{\boldsymbol{e}}_{R}$
direction, especially in the later stage. As mentioned previously,
the neoclassical orbit loss is just one of many loss mechanisms for
positrons. Since other loss mechanisms \cite{martin1999energy,0741-3335-35-SB-010,yoshino2000runaway,helander2000suppression,lehnen2008suppression,papp2011runaway,finken2007runaway}
are in general faster, the in-plasma annihilation probability observed
in experiments should be even less. Finally, because the orbit drift
does not depend on the electric charge, note that runaway electrons,
which will have their own signature (like visible damage to the wall),
will strike exactly on the same side of the vacuum vessel.

Early in 1986, Surko et al. proposed to diagnose the transport process
by injecting positrons into tokamaks \cite{surko1986use}. The annihilation
spectrum of positrons in thermal plasmas was also studied \cite{ziemelis1976positrons}.
Now it is clear that the large amounts of positrons produced by tokamaks
themselves can be used as a diagnostic tool as well. Since most of
the positrons are annihilated outside the plasma, the positron diagnostic
in tokamaks cannot detect plasma properties directly as in PET (Positron
Emission Tomography) \cite{PETbook}. However, positron diagnostic
is still possible using our knowledge of the dynamics of the runaway
positrons. The annihilation spectrum, which can be recorded by the
gamma spectrometer, is very characteristic and easily identifiable.
The intensity, breadth, and shift of the annihilation peak in the
spectrum reflect the properties of the plasma that determines the
runaway dynamics. Moreover, the time history of the positron annihilation
is also an important indication to some events in tokamaks, such as
a disruption or rf heating, which produce a burst of positron runaways
at a specific time. The annihilation locations on the first wall and
the emission directions of the gamma ray provide information about
the phase space coordinates of the runaway positrons at the end of
their journey, from which we can also infer the trajectories of runaway
electrons in the phase space. Theoretical and experimental studies on
these topics will be reported in future publications.
\begin{acknowledgments}
This research is supported by the JSPS-NRF-NSFC A3 Foresight Program
in the field of Plasma Physics (NSFC-11261140328), the National Natural
Science Foundation of China (NSFC-11305171), ITER-China Program (2014GB124005
and 2013GB111000), the Fundamental Research Funds for the Central
Universities (WK2030020022), and the CAS Program for Interdisciplinary
Collaboration Team. We thank Professor Tieshuan Fan and Di Hu for
fruitful discussion.
\end{acknowledgments}

\bibliographystyle{apsrev}

\end{document}